\documentclass[pre,aps,floats,superscriptaddress,floatfix,twocolumn,longbibliography]{revtex4-1}
\usepackage{amssymb,amsmath}
\usepackage{graphicx}
\usepackage{dcolumn}
\usepackage{bm}
\usepackage{comment}
\usepackage{hyperref}
\usepackage{makecell}
\usepackage{changepage}
\usepackage{multirow}
\usepackage[table]{xcolor}
\usepackage[mathscr]{euscript}
\usepackage{csquotes}

\begin{document}

\title{Ordering kinetics in the active Ising model}

\author{Sayam Bandyopadhyay}
\email{bandyosayam@gmail.com}
\affiliation{School of Mathematical \& Computational Sciences, Indian Association for the Cultivation of Science, Kolkata -- 700032, India.}

\author{Swarnajit Chatterjee}
\email{swarnajit.chatterjee@uni-saarland.de}
\affiliation{Center for Biophysics \& Department for Theoretical Physics,
Saarland University, 66123 Saarbr\"ucken, Germany}

\author{Aditya Kumar Dutta}
\affiliation{School of Mathematical \& Computational Sciences, Indian Association for the Cultivation of Science, Kolkata -- 700032, India.}

\author{Mintu Karmakar}
\affiliation{School of Mathematical \& Computational Sciences, Indian Association for the Cultivation of Science, Kolkata -- 700032, India.}

\author{Heiko Rieger}
\email{heiko.rieger@uni-saarland.de}
\affiliation{Center for Biophysics \& Department for Theoretical Physics,
Saarland University, 66123 Saarbr\"ucken, Germany}
\affiliation{INM – Leibniz Institute for New Materials, Campus D2 2, 66123 Saarbr\"ucken, Germany}

\author{Raja Paul}
\email{raja.paul@iacs.res.in}
\affiliation{School of Mathematical \& Computational Sciences, Indian Association for the Cultivation of Science, Kolkata -- 700032, India.}

\begin{abstract}
We undertake a numerical study of the ordering kinetics in the two-dimensional ($2d$) active Ising model (AIM), a discrete flocking model with a conserved density field coupled to a non-conserved magnetization field. We find that for a quench into the liquid-gas coexistence region and in the ordered liquid region, the characteristic length scale of both the density and magnetization domains follows the Lifshitz-Cahn-Allen (LCA) growth law: $R(t) \sim t^{1/2}$, consistent with the growth law of passive systems with scalar order parameter and non-conserved dynamics. The system morphology is analyzed with the two-point correlation function and its Fourier transform, the structure factor, which conforms to the well-known Porod’s law, a manifestation of the coarsening of compact domains with smooth boundaries. We also find the domain growth exponent unaffected by different noise strengths and self-propulsion velocities of the active particles. However, transverse diffusion is found to play the most significant role in the growth kinetics of the AIM. We extract the same growth exponent by solving the hydrodynamic equations of the AIM.
\end{abstract}

\maketitle

\section{Introduction}
\label{introduction}
Active matter systems involve the movement of large assemblies of individual active particles that consume energy to self-propel and exhibit collective behavior in a non-equilibrium steady state~\cite{ramaswamy,marchetti2013hydrodynamics,needleman2017active,gompper20202020}. Collective motion is ubiquitous in nature, observed in a wide array of different living systems over a range of scales, from macroscopic fields like fleets of birds~\cite{Ballerini} and schools of fish~\cite{Becco,Calovi} to microscopic scales like hoards of bacteria~\cite{Steager,Peruani}, cytoskeletal filaments and molecular motors~\cite{schaller2010polar,sumino2012large,sanchez2012spontaneous}. It leads to the emergence of ordered motion of large clusters, called flocks, with a typical size larger than an individual~\cite{ramaswamy,collectivemotion,shaebani2020computational,de2015introduction,krishnan2010rheology}. Over the past two decades, new models have emerged to understand the various physical principles governing active matter systems \cite{shaebani2020computational}.

Vicsek and collaborators years ago introduced a minimal model~\cite{Vicsek} of active particles that move with a constant speed and orient via a ferromagnetic interaction with a neighborhood similar to the XY model. In the Vicsek model (VM) activity can stabilize the ordered phase even in two dimensions which is not possible in the $2d$ XY model where the long-range fluctuation destroys the ordered phase following Mermin-Wagner theorem ~\cite{toner1995long,toner1998flocks}. Then, about a decade ago, Solon and Tailleur introduced the active Ising model (AIM)~\cite{AIM_solon_intro,solon2015flocking} where a discrete symmetry replaces the continuous rotational symmetry of the VM. In the AIM, each particle assumes two possible states allowing the particle to propel in a preferred direction which changes upon interaction with other particles at the same lattice site.  The AIM retains the essential part of the VM physics and exhibits flocking behavior with three different phases at steady state: disordered gas at high noise and low densities, polar liquid at low noise and high densities, and a phase-separated liquid-gas coexistence state at intermediate densities. However, a key difference between the VM and the AIM arises in the steady-state behavior of the coexistence region. In this region, AIM shows a {\it macrophase} separation associated with normal density fluctuations whereas the VM is characterized by a {\it microphase} separation with giant density fluctuations. The flocking transition in the AIM is a first-order liquid-gas phase transition similar to the VM, however, for zero activity, despite the dynamics being non-equilibrium, the AIM shows a second-order phase transition belonging to the Ising universality class. 

Although significant progress has been made to understand the steady state properties of various active systems~\cite{toner2005170,giomi2013defect,MIPS,solonVM,criticalABP2018,capillary2020adam,Raja_APM,chatterjee2020flocking,kursten2020dry,fruchart2021NR,solon2022susceptibility,chatterjee2022polar,codina2022small,TSVM2023,karmakar2023jamming}, there is much to explore in the realm of ordering kinetics in active systems that relaxes to a non-equilibrium steady state (NESS). Understanding the intrinsic non-equilibrium dynamics that drive an active system towards its steady state is of fundamental as well as practical relevance. Unlike active systems, ordering kinetics in non-equilibrium passive systems have been studied over several decades~\cite{alan_bray,bray1993theory,sanjay_puri,aging2012,puri2014rfim,kumar2017ordering,rbcm}. Domain growth in passive systems with non-conserved scalar order parameters follows the Lifshitz-Cahn-Allen (LCA) growth law: $R(t) \sim t^{1/2}$ (\textit{Model A} of order-parameter kinetics) whereas passive systems with conserved order parameter follow a  Lifshitz-Slyozov-Wagner (LSW) growth law: $R(t) \sim t^{1/3}$ (\textit{Model B} of order-parameter kinetics), $R(t)$ being average size of domains. However, domain growth in \textit{Model C} systems where a non-conserved scalar order parameter is coupled to a conserved density field (similar to the AIM), is governed by the growth exponent of either \textit{Model A} or \textit{Model B} where it depends on the quenching regime. Quench into the order-disorder coexistence region leads to a $R(t) \sim t^{1/3}$ growth whereas the asymptotic growth law for a quench into the ordered region is $R(t) \sim t^{1/2}$~\cite{chate2002modelC}.

Utilizing tools that quantify the ordering kinetics of passive systems, several active systems have been explored. These include Active Model B~\cite{wittkowski2014scalar,pattanayak2021AMB,pattanayak2021domain}, active nematics~\cite{mishra2014aspects}, self-propelled particles in disordered medium~\cite{das2018ordering}, Model B with nonreciprocal activity~\cite{saha2020scalar}, Kuramoto oscillators~\cite{rouzaire2022dynamics}, active Brownian particles~\cite{dittrich2023growth} and motility-induced phase separated (MIPS) clusters~\cite{caporusso2023dynamics}. Moreover, an interesting observation of multiple coarsening length scales was made in the prototypical VM~\cite{Coarsening2020VM} where velocities are found to align over a faster-growing length scale compared to density. Another intriguing result of an active system with a non-conserved vector order parameter following the growth law of the non-conserved scalar order parameter field has also been observed~\cite{Dikshit_2023}. Since AIM is a minimal flocking model with a rich phase behavior, studying the growth kinetics of this model will allow us to interpret the origin of large flocks in terms of microscopic interactions. 

In this paper, we explore the phase ordering kinetics of the AIM. Quenching the AIM inside the spinodal region results in the formation of small positively or negatively magnetized clusters which in the late stage of the coarsening merge to form a single, macroscopic domain of one spin type~\cite{solon2015flocking}. A few questions arise in this context: (a) Does the domain morphology follow the same pattern and growth law for quenches into the coexistence and in the ordered liquid region? (b) Since the AIM possesses both conserved and non-conserved order parameters, how does the AIM growth law relate to the established growth law of similar passive systems? (c) Do the density and magnetization correlate over the same length scale?~\cite{Coarsening2020VM} (d) What is the impact of noise and particle activity on the domain growth? and (e) What is the role of diffusion in the domain growth dynamics? We address these issues by analyzing the ordering dynamics of the $2d$ AIM on a square lattice via Monte Carlo simulations and by solving the AIM hydrodynamic equations using the finite difference method.

This paper is organized as follows. In Sec.~\ref{model} we discuss the model and then present the details of numerical simulations in Sec.~\ref{simulation}. In Sec.\ref{results}, we present the growth law of the AIM from both numerical simulation and hydrodynamic description. Finally, in Sec.~\ref{conclusion}, we conclude this paper with a summary and discussion of the results.

\section{Model}
\label{model}
We consider $N$ particles on a two-dimensional square lattice $L\times L$ with periodic boundary conditions. Thus the average particle density is  $\rho_{0} = N/L^2$. Each lattice site $i$ can accommodate an arbitrary number of particles $n_i^{\sigma}$ with spin $\sigma =\pm 1$. Defining local density $\rho_i=n_i^{+} + n_i^{-}$ and magnetization $m_i = n_i^+ - n_i^-$, we note that $\rho_i$ has no upper bound, while $m_i$ is bounded by $\rho_i$: $-\rho_i \leq m_i \leq \rho_i$.  Each particle with a given spin state $\sigma$ can either flip to $-\sigma$ or jump to a nearest-neighbor site probabilistically.

The flipping rates are derived from a local ferromagnetic Ising Hamiltonian defined as~\cite{AIM_solon_intro,solon2015flocking}:
\begin{equation}\label{hamiltonian}
    H_i = - \frac{J}{2\rho_i} \sum_j \sum_{k\ne j} \sigma_i^j \sigma_i^k \, ,
\end{equation}
where $J$ is the coupling between any two particles at site $i$. Local interaction implies that a particle can align with the average direction of all other particles at the same site. The Hamiltonian of Eq.~\eqref{hamiltonian} can be rewritten as:
\begin{equation}
H_i = - \frac{J}{2} \left( \frac{m_i^2}{\rho_i} - 1 \right) \, .
\end{equation}
When a particle in spin-state $\sigma$ flips, $m_i$ changes to $m_i -2 \sigma$. The energy difference is then:
\begin{equation}
\Delta H_i = \frac{2J}{\rho_i} \left(\sigma m_i - 1\right) \, .
\end{equation}
A particle with spin $\sigma$ then flips its state according to the transition rate:
\begin{equation}\label{flip}
    W_{\rm flip}(\sigma \rightarrow - \sigma) = \gamma \exp\left[-\frac{2\beta J}{\rho_i}\left(\sigma m_i - 1\right)\right] \, ,
\end{equation}
where $\gamma$ is the rate of particle flipping when $\sigma m_i = 1$. The transition rate is chosen to fulfill the detailed balance without hopping with respect to the Hamiltonian $H_i$. In this paper, we choose $J=1$, and $\gamma=1$ without any loss of generality. The parameter $\beta$, denoted as ``inverse temperature'', $\beta=T^{-1}$, in passive systems, controls the flip noise strength. Although the system under consideration is athermal, we denote the parameter $T$ as ``temperature'' from now on. 

Moreover, each particle performs a biased diffusion on the lattice depending on the spin state $\sigma$. Particles perform a one-dimensional biased hopping with self-propulsion $\epsilon$ along a direction ${\bf p}$ via the rate~\cite{solon2015flocking}:
\begin{equation}\label{hop}
    W_{\rm hop}(\sigma,{\bf p}) = D(1+\sigma \epsilon {\bf p} \cdot {\bf e_x}) \, .
\end{equation}
The presence of other particles does not influence hopping rates and hence is independent of particle density. The hopping rate $D=1$ is constant along the upward and downward directions ($\pm y$-direction). The parameter $\epsilon \in [0, 1]$ controls the asymmetry between the purely diffusive limit $\epsilon = 0$ and the purely ballistic limit $\epsilon = 1$, while $D$ controls the overall hopping rate. On average, a particle drifts with speed $v = 2D\epsilon$ in the direction set by the sign of its spin state (where lattice spacing is 1), while the total hopping rate $4D$, remains constant.

\section{Simulation details}
\label{simulation}
A Monte Carlo (MC) simulation of the stochastic process defined above evolves in unit Monte Carlo steps (MCS) $\Delta t$ resulting from a microscopic time $\Delta t/N$. During $\Delta t/N$, a randomly chosen particle with spin $\sigma$ flips with probability $W_{\rm flip}\Delta t$ or hops to one of the neighboring sites with probability $W_{\rm hop}\Delta t$. Consequently, $1 - [4D + W_{\rm flip}]\Delta t$ is the probability that the particle does nothing, and minimizing this we obtain $\Delta t=[4D + \exp(2\beta)]^{-1}$.

To study the morphology of the system during phase ordering kinetics, we use the two-point equal-time $(t)$ correlation function of the local scalar order parameters. The notion utilizes the spatial fluctuations in the density and magnetization fields to estimate~\cite{Dikshit_2023}:
\begin{equation}\label{correlation_density}
    C_{\rho\rho}(r,t) =  \frac{1}{L^2} \sum_{i=1}^{L^2} \langle \Delta \rho_{i,t} \Delta \rho_{i+r,t} \rangle 
\end{equation}  
\begin{equation}\label{correlation_mag}
    C_{ mm}(r,t) =  \frac{1}{L^2} \sum_{i=1}^{L^2}  \langle \Delta m_{i,t} \Delta m_{i+r,t} \rangle \, ,
\end{equation}
where $\langle \cdots \rangle$ denotes averaging over independent initial realizations, $\Delta \rho_{i,t} = \rho_i - \rho_0$ and $\Delta m_{i,t}=m_i - m_0$ are the local fluctuations in density and magnetization from the mean, respectively. The above definition of $C_{\rho\rho}$ characterizes the morphology of the spatial structures and $C_{ mm}$ evaluates the correlations in polar alignment between evolving structures separated by distance $r$. Since we observe that $C_{\rho\rho}$ and $C_{mm}$ behave similarly in the AIM (see below), we focus here on $C_{\rho\rho}$, which we denote as $C$ [$C_{\rho\rho}(r,t) \equiv C(r,t)$] from now on. Following a temperature quench from a random initial configuration into the ordered state, clusters of both spin types appear and grow with time. Similar morphology of the evolving domains with average domain size $R(t)$ would correspond to a dynamical scaling relation \cite{sanjay_puri,alan_bray,bray1993theory}: 
\begin{equation}\label{correlation}
    C(r,t) =  f \left (\frac{r}{R(t)} \right)
\end{equation}
where $f(x)$ is a time-independent scaling function. $R:q(t)$, estimated from the decay of $C(r,t)$ generally show a power-law growth \cite{sanjay_puri,alan_bray,bray1993theory}:
\begin{equation}\label{lengthscale}
    R(t) \sim t^\theta \, ,
\end{equation}
with $\theta$ as the growth exponent. Typically, the morphology of an ordering system is studied by scattering experiments, which measure the structure factor $S(k,t)$, defined by the Fourier transform of the correlation function $C(r,t)$:
\begin{equation} \label{sf}
    S(k,t)=\int_{-\infty}^{\infty} C(r,t) e^{ikr} dr \, ,
\end{equation}
and has a dynamical scaling form in $d$ dimensions:
\begin{equation}\label{sfscaling}
    S(k,t) =  R(t)^{d} g\left[ kR(t) \right] \, .
\end{equation}
For scalar order parameters like the density field, the short-distance (large-$k$) behavior of the structure factor scaling function is given by Porod's law (for domains with smooth boundaries or scattering off sharp domain interfaces) which corresponds to $g(k) \sim k^{-(d+1)}$. 
Next, we present results for model parameters set by the average particle density $\rho_0$, temperature $T=\beta^{-1}$, self-propulsion velocity $\epsilon$, and diffusion constant $D$. 

\section{Results}
\label{results}


\subsection{Phase diagram and domain morphology}
\begin{figure}[t]
\centering
\includegraphics[width=\columnwidth]{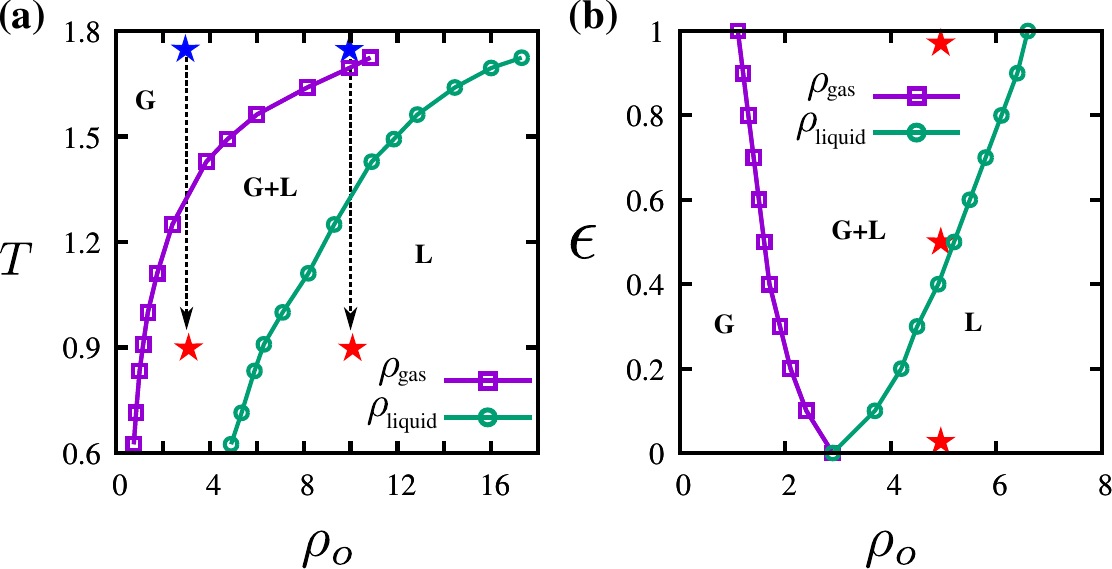}
\caption{(Color online) Phase diagrams of the AIM showing the quench directions at fixed densities. (a) $(T,\rho_0)$ phase diagram for self-propulsion velocity $\epsilon=1$. (b) $(\epsilon,\rho_0)$ phase diagram at fixed temperature $T=0.9$. G and L denote the disordered gas and polar liquid regions whereas G+L denotes the phase-coexistence region. The (blue) stars in (a) indicate the initial high-temperature points in the phase diagram from where the system is quenched (indicated by arrows) into the coexistence region ($\rho_0=3$) and to the liquid region ($\rho_0=10$) [(red) stars at lower temperature]. (Red) stars in (b) mark the quench points in the $(\epsilon,\rho_0)$ plane. $\rho_{\rm gas}$ (open square, delimits G and G+L) and $\rho_{\rm liquid}$ (open circle, delimits G+L and L) are the gas and liquid binodal, respectively.}
\label{fig:phase_diagram}
\end{figure}

The steady-state behavior of the AIM is summarized in the temperature-density ($T - \rho_0$) [Fig.~\ref{fig:phase_diagram}(a)] and velocity-density ($\epsilon - \rho_0$) [Fig.~\ref{fig:phase_diagram}(b)] phase diagrams. The general structure of the phase diagrams consists of a gas phase (G), a liquid phase (L), and a liquid-gas coexistence region (G+L) separated by the gas and liquid binodals. Qualitatively similar results were obtained earlier in Ref.~\cite{solon2015flocking}. However, we obtain a critical temperature $T_c \simeq 2$ (above which no phase separation occurs regardless of the density, the critical density $\rho_c=\infty$) twice as large as in Ref.~\cite{solon2015flocking}. This can be understood if one compares the flipping rate of Ref.~\cite{solon2015flocking}, $W_{\rm flip}(\sigma \rightarrow - \sigma) = \gamma \exp\left(-\sigma \beta \frac{m_i}{\rho_i}\right)$ with Eq.~\eqref{flip} for $J=1$. This indicates that the effective $\beta$ considered in this paper is approximately twice as large as the $\beta$ value in Ref.~\cite{solon2015flocking}. In the $\epsilon - \rho_0$ phase diagram, the two binodals converge at a critical density $(\rho_c \simeq 2.9)$ for vanishing self-propulsion ($\epsilon=0$) that signifies a second-order phase transition belonging to the Ising universality class~\cite{AIM_solon_intro}.

We chose to quench the random initial systems in two different regimes of the temperature-density phase diagram shown by the arrows in Fig.~\ref{fig:phase_diagram}(a). The quench occurs instantaneously from a very high-temperature regime ($T\rightarrow \infty$) shown by the blue stars to the liquid-gas coexistence region or the polar liquid region, denoted by red stars, with the same final temperature $T=0.9$ ($T<T_c$, corresponds to $\beta=1.1$) but different densities.

To quantify the coarsening dynamics, we performed simulations on a square lattice of size $400^2$ with periodic boundary conditions applied on both sides. Following quench at time $t=0$, the system evolves up to $t=10^5$ using the MC algorithm described in Sec.~\ref{simulation}. All numerical data presented here are averaged over at least 300 independent realizations.

\begin{figure*}[!t]
\centering
    \includegraphics[width=\textwidth]{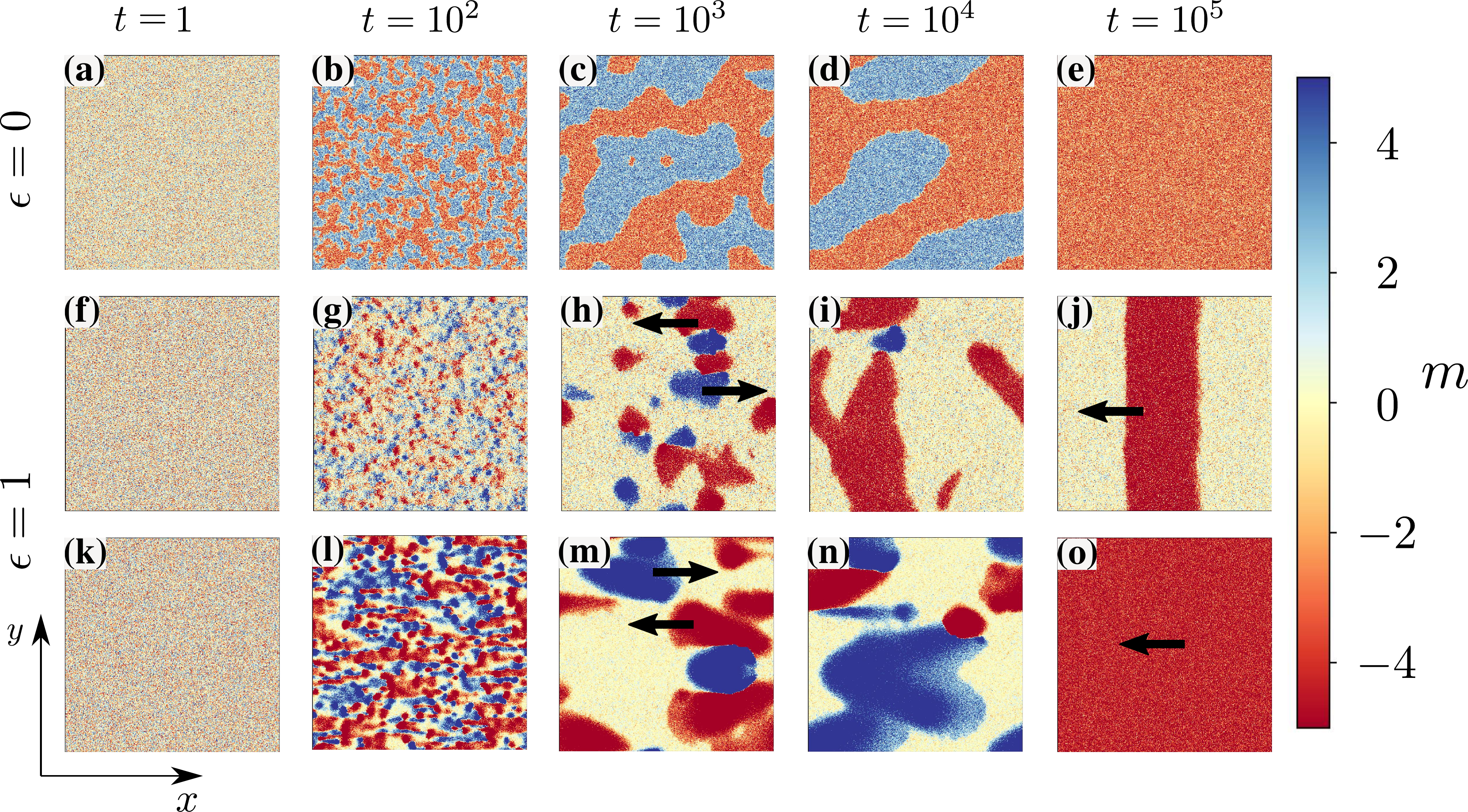}
    \caption{(Color online) Time evolution snapshots of the local magnetization field on a $400^2$ system after a quench from $T=\infty$ to $T=0.9$ for $\epsilon=0$ (top panel) and $\epsilon=1$ (middle and lower panel). The color bar denotes magnetization per site. (a--e) Curvature-driven domain growth of the diffusive Ising model ($\rho_0=5$, $\epsilon=0$). (f--j) Growth dominated by the dynamics of spinodal decomposition after the system is quenched inside the spinodal region ($\rho_0=3$, $\epsilon=1$) and (k--o) domain growth mediated by the merging of high-density clusters of $\sigma=\pm 1$ after the system is quenched deep inside the ordered liquid regime ($\rho_0=10$, $\epsilon=1$). Black arrows denote the direction of the movement of the clusters and bands.}
    \label{fig:dynamics_snap}
\end{figure*}

A typical simulation of coarsening dynamics starts with a homogeneous initial configuration where particles with $\sigma=\pm 1$ are distributed randomly on each lattice site. This can also be described as an equilibrium configuration at infinite temperature because all configurations are equally likely. Then, after a quench well below the critical temperature ($T_c \simeq 2$), the homogeneous initial configuration starts evolving in time, and subsequent dynamics are governed by the formation and growth of $\sigma=1$ and $\sigma=-1$ rich domains. Such a time evolution of the local magnetization field is shown in Fig.~\ref{fig:dynamics_snap} for $\epsilon=0$ and $\epsilon=1$. In the latter scenario, time evolution has been shown for a quench in both the coexistence (middle row) and the liquid regions (bottom row). The non-equilibrium steady states (NESS) are characterized by a single band (at lower density) and a polar liquid (at higher density).

The top panel, Fig.~\ref{fig:dynamics_snap}(a--e), depicts the ordering kinetics of a purely diffusive $(\epsilon=0)$ polar liquid at $\rho_0=5$. The domain morphology with increasing time exhibits a close resemblance to the passive Ising model~\cite{2dIsing_growth}. In the Ising model, the driving force for domain growth is the curvature of the domain wall, since the system surface energy can only decrease through a reduction in the total net surface area. In the $\epsilon=0$ limit, particles do not form high-density domains due to the diffusive movement of particles. Therefore, density-wise, the system remains homogeneous as we observe a steady growth of small, high-curvature to large, low-curvature domains. This is unsurprising as the $2d$ AIM for $\epsilon=0$ belongs to the same universality class of the passive $2d$ Ising model~\cite{solon2015flocking}. 

Next, we looked at the evolution of AIM with self-propulsion velocity ($\epsilon=1$)  quenching the system inside the spinodal  [Fig.~\ref{fig:dynamics_snap}(f--j)] and homogeneous ordered regions [Fig.~\ref{fig:dynamics_snap}(k--o)]. The average densities representing these two regions correspond to $\rho_0=3$ and 10, respectively. Inside the spinodal region, the growth dynamics is driven by spinodal decomposition and result in the formation of numerous small clusters of negative and positive spins $(t=10^2)$. The coarsening then stems from the merging of these clusters $(t=10^3$ and $t=10^4)$, until a single macroscopic domain emerges $(t=10^5)$ in the steady state. A quench deep inside the homogeneous ordered region also results in similar dynamics of cluster formation and coalescence of those clusters into a single large liquid domain (in this paper, for $\epsilon>0$, the word cluster is used interchangeably with domain). With extreme self-propelled particles ($\epsilon=1$), although the high-density clusters are strongly biased along the horizontal directions, they can also grow along the transverse direction due to the constant transverse diffusion $D$. When these clusters merge into a larger cluster, it always tries to minimize the surface energy by decreasing the surface area. Accordingly, the coarsening process leads to a single band [Fig.~\ref{fig:dynamics_snap}(j)] with domain walls having the lowest curvature (the curvature of a straight line is zero, but a liquid band with a perfectly straight domain boundary can only happen when there is no thermal fluctuation). The focus of this study is therefore to analyze the coarsening dynamics shown in Fig.~\ref{fig:dynamics_snap} which we will do next.


\subsection{Dynamical scaling}
\begin{figure*}[!t]
    \includegraphics[width=0.7\textwidth]{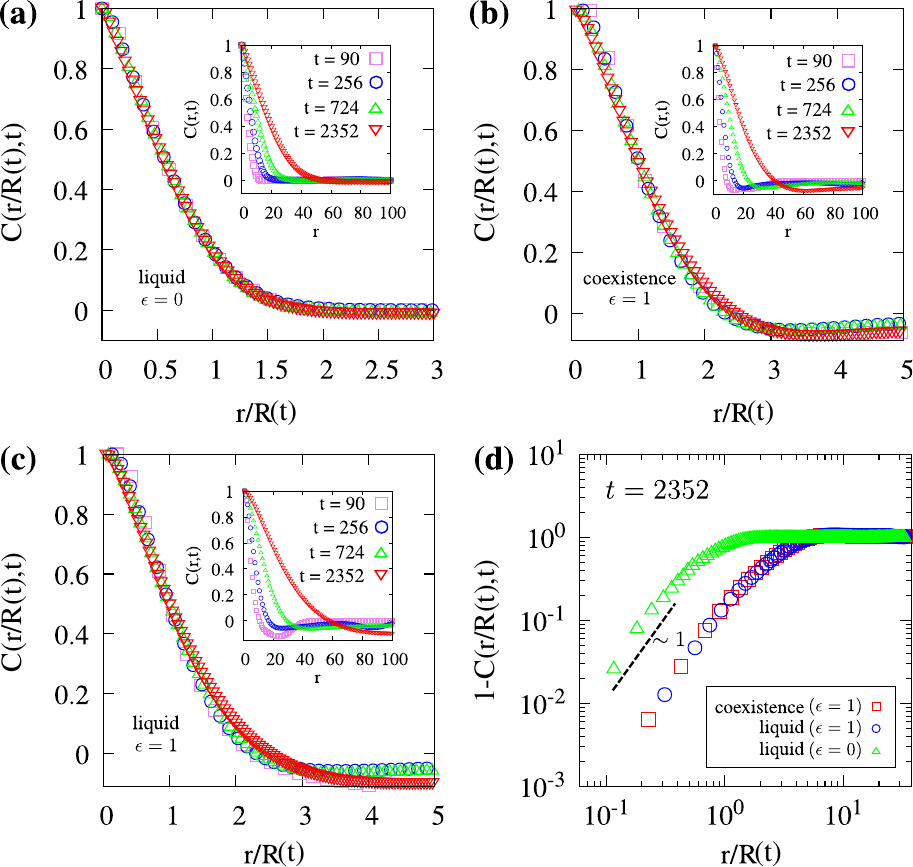}
    \caption{(Color online) (a--c) Scaled correlation functions $C(r/R(t),t)$ versus $r/R(t)$, for the evolution of the $2d$ AIM shown in Fig.~\ref{fig:dynamics_snap}. Unscaled versions are shown in the inset. ``Liquid'' and ``coexistence'' signify the region where the system is quenched, $\epsilon$ denotes particle speed. (d) Log-log plot of $1-C(r/R(t))$ versus $r/R(t)$. The cusp exponent is estimated as $\alpha \sim 1$. The parameters correspond to Fig.~\ref{fig:dynamics_snap}.}
    \label{fig:correlation_plot}
\end{figure*}

\begin{figure*}[!t]
    \includegraphics[width=\textwidth]{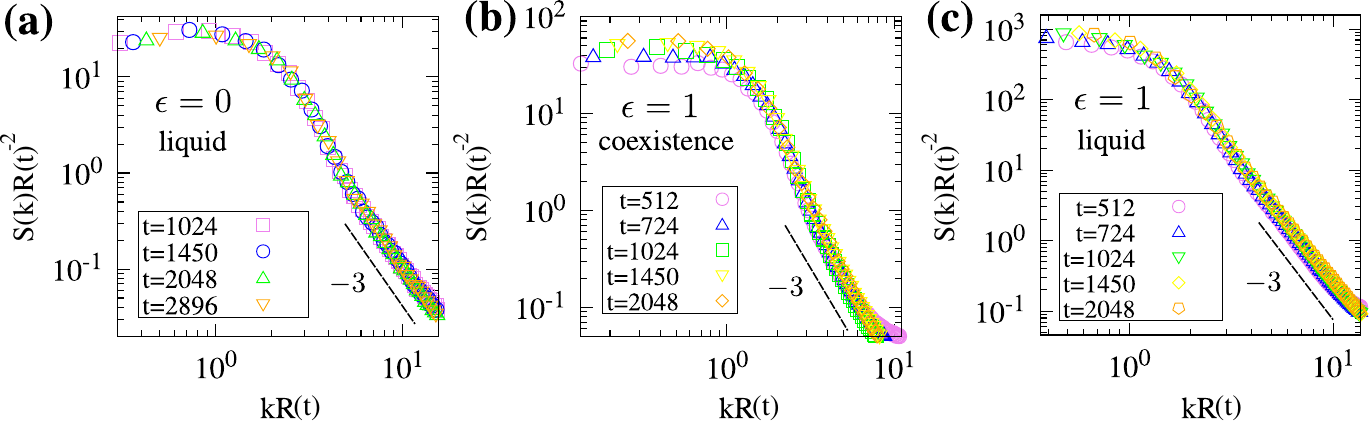}
    \caption{(Color online) (a--c) Scaled structure factors, $S(k,t)R(t)^{-2}$ versus $kR$, for the Fourier transform of the correlation function data sets corresponding to the same values of time.  The line of slope $\simeq -3$ denotes the Porod's law: $S(k)\sim k^{-(d+1)}$ for $d=2$.}
    \label{fig:struc_plot}
\end{figure*}
In the theory of phase-ordering kinetics, the scaling hypothesis states that if the system is characterized by a single length scale $R(t)$ [$R(t)$ is equivalent to the average domain size], the domain morphology is statistically the same at all times, apart from a scale factor. When all domain lengths are measured in units of $R(t)$, the equal-time pair correlation function should exhibit the dynamical scaling property of Eq.~\eqref{correlation}. The equal-time correlation function is a non-equilibrium quantity so as $R(t)$ which can be estimated from the decay of the correlation function. To identify the self-similar behavior of the evolving domains we plot the correlation function in Fig.~\ref{fig:correlation_plot}. The scaled $C(r/R(t),t)$ and unscaled $C(r,t)$ correlation functions are shown in Fig.~\ref{fig:correlation_plot}(a--c) for  $\epsilon=0$ and $\epsilon=1$. $R(t)$ is determined from the distance over which the correlation function decays to e.g., 0.2 of its maximum value, that is, $C(r,t)=0.2C(0,t)$. 

As the system coarsens, the correlation function decays slowly [insets of Fig.~\ref{fig:correlation_plot}(a--c)], signifying the growth of the characteristic length scale $R(t)$. Upon rescaling the spatial coordinates by this length scale, the correlation function at different times collapses onto a single function $C(r/R,t)$ as shown in Fig.~\ref{fig:correlation_plot}(a--c), thus confirming a universal coarsening behavior with time. Such scaling behavior implies that the structure is time-invariant and consistent with a power-law growth of $R(t)$ with increasing $t$. This scaling hypothesis is satisfied when the growing length is much smaller than the system size to avoid the finite size effect. We further examine the AIM domain morphology by approximating the small distance behavior of the scaled two-point correlation function 
\begin{equation}\label{cusp}
  1-C(r) = \Bar{C}(r) \sim r^\alpha 
\end{equation}
 in Fig.~\ref{fig:correlation_plot}(d) which yields the cusp exponent $\alpha \sim 1$. This signifies the existence of sharp domain interfaces [see Fig.~\ref{fig:dynamics_snap}(h) and Fig.~\ref{fig:dynamics_snap}(m)] and translates into the power-law behavior of the scaled structure factor plotted in Fig.~\ref{fig:struc_plot}~\cite{puri2014rfim}. Fig.~\ref{fig:correlation_plot}(d) also signifies that the domain structure of the AIM for $\epsilon=1$ is statistically self-similar for quenches into the coexistence and liquid region whereas different from the domain morphology of the AIM for $\epsilon=0$, as evident from Fig.~\ref{fig:dynamics_snap}. 

In Fig.~\ref{fig:struc_plot}, we plot the scaled structure factor, $S(k,t)R(t)^{-2}$ versus $kR(t)$, which is the Fourier transform of the correlation function. In Fourier space, Eq.~\eqref{cusp} translates into the following power-law behavior of the structure factor: $S(k)\sim k^{-(d+\alpha)}$ and therefore, the large-$k$ behavior of the structure factor tail generates a slope $\sim -3$ (in log-log plot) for $d=2$ and $\alpha=1$ which denotes ``Porod’s decay'': $S(k)\sim k^{-(d+1)}$, associated with scattering from sharp interfaces~\cite{alan_bray,bray1993theory}. This naturally originates from the long-range ordering in AIM leading to compact high-density clusters with smooth boundaries. Domains with rough morphologies having fractal interfaces do not follow Porod's law and the large-$k$ tail of the scaled structure factor yields a non-integer exponent~\cite{chatterjee2018clock}. A similar violation of the Porod law was also observed in the coarsening of the VM due to the irregular morphology associated with the cluster boundaries \cite{Supravat2012spatstructGNF,Coarsening2020VM}. $C(r,t)$ [and consequently, $S(k,t)$] exhibits two distinct power laws for small and large $r/R(t)$ limits [large and small $kR(t)$ limits] in the VM~\cite{Supravat2012spatstructGNF} which we do not observe in the AIM. Furthermore, density fluctuations might also play a role in determining whether a system will follow or violate Porod’s law. In the VM, giant density fluctuations break large liquid domains and restrict the formation of large compact domains (which eventually manifest in the microphase separation of the coexistence region in the steady-state)~\cite{solonVM} and might be responsible for the non-Porod behavior of the system~\cite{Supravat2012spatstructGNF}. On the other hand, AIM obeys the Porod behavior where density fluctuations are normal in the liquid phase~\cite{solon2015flocking} and the steady-state manifests a bulk phase separation.


\subsection{Growth law}
\begin{figure*}[!t]
    \centering
    \includegraphics[width=\textwidth]{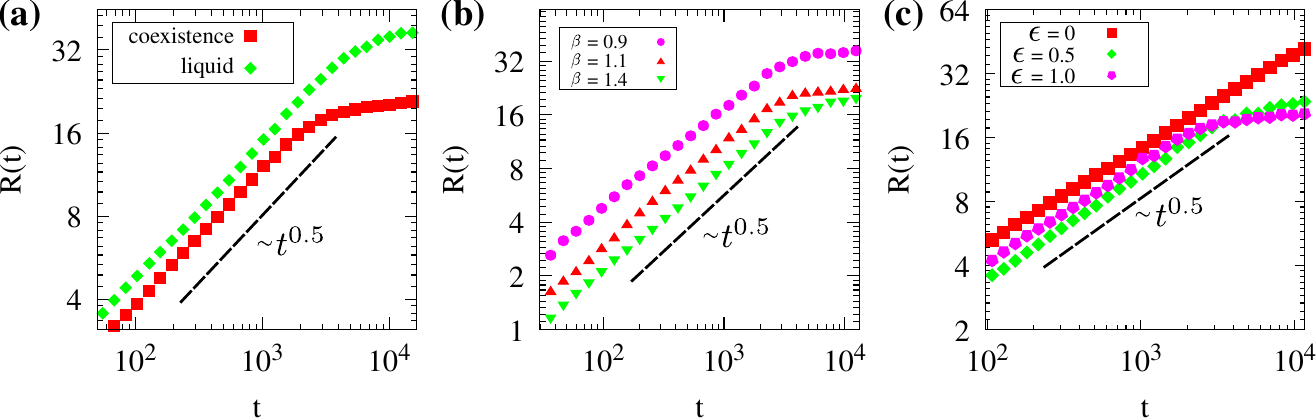}
    \caption{(Color online) (a) $R(t) \sim t^{1/2}$ for a quench into the coexistence (solid red square) and liquid (solid green diamond) regimes. The late time saturation of $R(t)$ signifies that the system has reached the corresponding NESS. As the density is higher for the liquid quench, the saturation appears late. Parameters: $\rho_0=3$ (coexistence), $\rho=10$ (liquid), $L=400$, $\beta=1.1$ and $\epsilon=1$. (b) $R(t)$ versus $t$ for different thermal noises [$T=\beta^{-1}$ with $\beta = 0.9$ (solid pink circle), $1.1$ (solid red triangle) and $1.4$ (solid green inverted triangle)] in the coexistence regime. Parameters: $L=400$, $\rho_0=3$ and $\epsilon=1$. (c) Role of self-propulsion on the characteristic length $R(t)$ as a function of $t$. For a larger $\epsilon$ [$=0.5$ (solid green diamond) and $1.0$ (solid pink inverted pentagon)], the system reaches the NESS faster while for $\epsilon=0$ (red square), the corresponding dynamics is slow. Parameters: $L=400$, $\beta=1.1$, and $\rho_0=5$. The growth law $R(t)\sim t^{1/2}$ is unaffected by thermal fluctuations and particle activity.}
    \label{fig:growth_plot}
\end{figure*}
In passive systems with non-conserved scalar order parameters, the late-stage domain growth is governed by the {\it diffusive} Lifshitz-Cahn-Allen (LCA) growth law $R(t) \sim t^{1/2}$~\cite{bray1993theory}. For non-conserved systems described by scalar fields such as the Ising model, the growth process is driven by the diffusion of the domain walls (the simplest form of topological defect) caused by the local changes in the order parameter. Diffusion also influences coarsening in non-conserved systems with vector fields, such as the $2d$ XY model, where domain evolution occurs when point topological defects, such as vortices and anti-vortices, diffuse, interact, and annihilate~\cite{yurke1993coarsening}. Both systems exhibit a diffusive growth exponent $\theta=\frac{1}{2}$ (in the XY model, $R(t) \sim (t/\ln t)^{1/2}$, the logarithmic correction is due to the free vortices~\cite{yurke1993coarsening}). Therefore, a 0.5 growth exponent signifies a coarsening process dominated by the diffusion of defects. For example, if the time evolution of a non-conserved scalar order parameter such as the magnetization $m$ follows the diffusive equation $\Dot{m} = D\nabla^2 m$, then the length scale will exhibit a $\sqrt{t}$ dependence.

However, as outlined in the Introduction, AIM dynamics is similar to the Model C dynamics~\cite{hohenberg1977} where a non-conserved magnetization field is coupled to a conserved density field. In Model C, the growth law for a quench into the order-disorder phase coexisting region appears to be $R(t) \sim t^{1/3}$ (reminiscent of Model B's growth law with a conserved order parameter~\cite{sanjay_puri}). Conversely, for a quench into the ordered region, the growth law follows $R(t) \sim t^{1/2}$~\cite{chate2002modelC} (reflecting the growth law of Model A with a non-conserved scalar order parameter).

Therefore, the immediate question we ask is whether the late-stage growth of the coarsening length scale in the AIM (an {\it active} system) follows the Model C growth laws, which vary depending on the quenching regime. To characterize this, we plot the length scale data with time in Fig.~\ref{fig:growth_plot} for various control parameters and quench regimes. We find that the late-stage growth kinetics of the domains in the coexistence and liquid regimes exhibit a $R(t) \sim t^{1/2}$ growth law [Fig.~\ref{fig:growth_plot}(a)]. We extract the same growth law for different temperatures [Fig.~\ref{fig:growth_plot}(b)], and self-propulsion velocities  [Fig.~\ref{fig:growth_plot}(c)]. Domains identified by the correlation of density and magnetization fields also show similar growth behavior (see Appendix~\ref{appA}). The domains exhibit larger sizes when quenching into the liquid regime compared to those formed when quenching into the coexistence region, primarily due to higher density. However, despite this difference, the coarsening length scale follows the same growth pattern in both scenarios [see Fig.~\ref{fig:dynamics_snap} at  $t=10^3$ and Fig.\ref{fig:growth_plot}(a)]. Also, notice that a larger $\beta$ in Fig.~\ref{fig:growth_plot}(b) signifies a reduced thermal noise that slows down the local ordering of spins. Thus smaller $\beta$ allows the formation of larger domains. Nevertheless, different thermal fluctuations exhibit the same growth as thermal noise is asymptotically irrelevant for ordering in systems that are free from disorder~\cite{sanjay_puri}.

Fig.~\ref{fig:growth_plot}(c) shows the increasing length scale with time for different self-propulsion velocity $\epsilon$. For the non-motile AIM ($\epsilon=0$), the system is purely diffusive, and domain growth proceeds through the coarsening of connected domains of $\sigma=\pm 1$ (curvature-driven growth facilitated by diffusing domain wall) [see Fig.~\ref{fig:dynamics_snap}(c--d)]. Therefore the domain morphology of the magnetization field looks very similar to the evolution of an Ising ferromagnet quenched below the critical temperature. Thus, a $\sim t^{1/2}$ growth law for $\epsilon=0$ similar to the pure Ising model is physically plausible. Furthermore, in the case of the non-motile AIM, we observe the same growth exponent $\theta=1/2$ for the conserved density field. We hypothesize that this similarity arises from the fact that the $\epsilon=0$ limit of the AIM belongs to the same universality class as the Ising model~\cite {solon2015flocking}. It is important to note that while AIM and Model C share the same combination of a non-conserved magnetization with Ising symmetry coupled to a conserved density, the crucial distinction lies in the fulfillment of detailed balance. Model C satisfies detailed balance, whereas even at $\epsilon=0$, AIM is not in equilibrium and fails to satisfy detailed balance regarding any distribution. Consequently, Scandolo et al.~\cite{scandolo2023active} argued that the critical point of AIM belongs to a distinct universality class from that of Model C. Moreover, it was suggested~\cite{scandolo2023active} that if a general AIM model is formulated where detailed balance gets restored in the zero self-propulsion limit, it would belong to the Model C universality class. This has recently been shown in a {\it thermodynamically consistent} model of the AIM~\cite{fodor2024aim}.

However, for $\epsilon>0$, the domains of each spin no longer remain connected and form high-density clusters that self-propel along the horizontal direction. As time progresses, these clusters spread in the transverse direction due to constant diffusion ($D=1$) and merge with other clusters. Therefore, the domain growth for $\epsilon>0$ is again a diffusive phenomenon, and consequently, the growth kinetics exhibits a growth law $R(t) \sim t^{1/2}$ similar to $\epsilon=0$. We have identified this novel mechanism of diffusion-driven domain growth in the AIM after a thorough investigation of the altered diffusion coefficient in Sec.~\ref{diffusion} and Appendix~\ref{appB}. 

\begin{figure}[!t]
    \centering
    \includegraphics[width=\columnwidth]{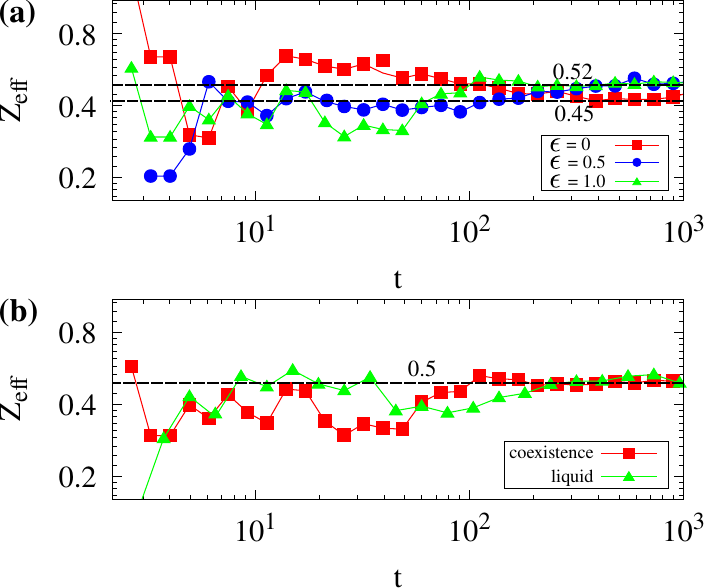}
    \caption{(Color online) Effective growth exponent $Z_{\rm eff}$ versus time $t$ for (a) different self-propulsion [$\epsilon=0$ (solid red square), $0.5$ (solid blue circle), and $1$ (solid green triangle), $\beta=1.1$ and $\rho_0=3$] and (b) quenching into different NESS [$\rho_0=3$ for coexistence (solid red square) and $\rho_0=10$ for liquid (solid green triangle), $\beta=1.1$ and $\epsilon=1$]. The dashed lines are a guide to the eyes.}
    \label{fig:zeff_plot}
\end{figure}

It should also be noted that while approaching the NESS via coarsening, the high-density AIM domains of individual spins, upon merging, try to minimize the surface energy by decreasing the surface area similar to the ordering dynamics of the passive Ising model. Therefore, although the system is active, the critical mechanisms (diffusion-dominated growth and minimization of surface energy during coarsening) of domain coarsening in AIM are similar to the $2d$ Ising model, and thus, it is not surprising that we extract a $R(t) \sim t^{1/2}$ growth law for both the passive and active models. For a more precise quantification of the asymptotic growth law, we determine the effective growth exponent, defined as:
\begin{equation}\label{zeff}
    Z_{\rm eff} = \frac{d [\ln R(t)]}{d[\ln t]} \, .
\end{equation}
In Fig.~\ref{fig:zeff_plot}(a), we plot $Z_{\rm eff}$ versus $t$ for different $\epsilon$ corresponding to the data in Fig.\ref{fig:growth_plot}(c). All data show extended flat regimes at late times (after $\sim t>10^2$). The effective exponent turns out to be in the regime $0.45<Z_{\rm eff}<0.52$, denoted by the dashed lines. We perform a similar study for the quench into the coexistence and liquid regimes. Fig.~\ref{fig:zeff_plot}(b) shows $Z_{\rm eff}$ vs. $t$ once again confirms $Z_{\rm eff}\sim 0.5$, irrespective of the quench regimes. 

Now, we aim to highlight the differences between the coarsening dynamics in the AIM and the VM, where both systems involve the coupling of a conserved density field with a non-conserved magnetization field. The main difference between the coarsening dynamics of AIM and VM lies in their growth laws. In the VM, the coarsening length scale of the density domains grows as $R(t) \sim t^{0.25}$, while the magnetization length scale grows much faster, $R(t) \sim t^{0.83}$~\cite{Coarsening2020VM}. The slower growth of the density domains can be attributed to the fractal morphology of the density field. Although density clusters in the VM are not compact, due to activity, individual particles may move between these clusters, transmitting the orientation order over larger length scales than the sizes of the density clusters, resulting in faster growth in the coarsening length scale of the magnetization~\cite{Coarsening2020VM}. On the other hand, in the AIM, both the density and magnetization fields exhibit a $R(t) \sim t^{0.5}$ growth law. The main ingredient responsible for this growth in the AIM is diffusion along the transverse direction which will be discussed in the following section. Regarding the morphology of the density and magnetization domains in these two models, the density fluctuation appears to be crucial. In the VM, density fluctuation is giant~\cite{solonVM}, whereas in the AIM, density fluctuation is normal~\cite{solon2015flocking}. Giant density fluctuations break large liquid domains, arrest band coarsening, and restrict the formation of large compact domains in the VM. In contrast, we observe compact high-density clusters with smooth boundaries in the AIM. This difference elucidates why we observe a faster growth in the density domains compared to the VM. The behavior of the magnetization fluctuations in the VM and AIM mirrors that of the density fluctuations: the VM exhibits giant fluctuations, while the AIM shows normal fluctuations. Despite this difference, both VM and AIM exhibit long-range order (LRO). The LRO in AIM is a natural consequence of its discrete symmetry~\cite{solon2015flocking}, while the continuous symmetry VM exhibits LRO~\cite{toner1995long} due to the non-equilibrium activeness of the particles. Therefore, the faster growth in the magnetization length scale in the VM (compared to AIM) can be attributed to activity, as discussed in Sec.~\ref{hydro}.


\subsection{Role of transverse diffusion}
\label{diffusion}
\begin{figure*}[!t]
    \centering
    \includegraphics[width=0.7\textwidth]{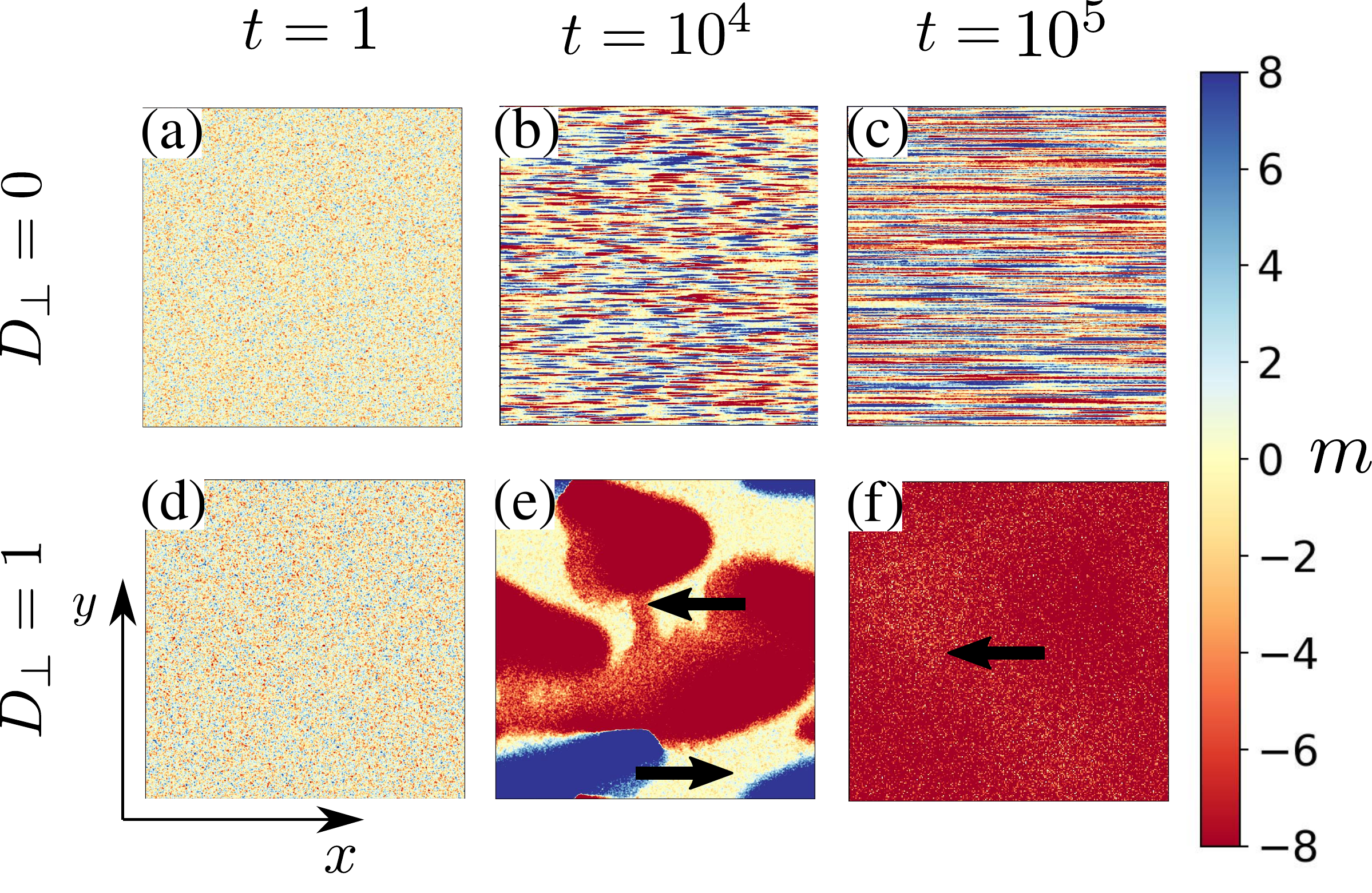}
    \caption{(Color online) Time evolution of the local magnetization field for $D_{\perp}=0$ [top panel, (a--c)] and $D_{\perp}=1$ [bottom panel, (d--f)] after quenching the system from a disordered gaseous phase to a high-density region. Parameters: $L=400, \beta=1.1$, $\rho_0=10$, and $\epsilon=1$.}
    \label{fig:D_zero_snap}
\end{figure*}
In the AIM, particles self-propel only in the horizontal direction with average velocity $2D\epsilon$. The directional hopping of the particles is a function of the spin type ($+\sigma$ or $-\sigma$). To test the hypothesis that the late time domain growth in the AIM is also a diffusion-driven process, we decompose the diffusion into two components, $D_{\perp}$ (along $\pm y$ direction) and $D_{\parallel}$ (along $\pm x$ direction). We vary $D_{\perp}$, keeping $D_{\parallel}=D=1$ henceforth, and show the domain evolution in Fig.\ref{fig:D_zero_snap}. 

The time evolution of the domains for $D_{\perp}=0$ is shown in Fig.~\ref{fig:D_zero_snap}(a--c). Starting from a disordered configuration, the time evolution of the system progresses via the formation of $1d$ rings with domains of alternating polarity along the horizontal direction ($x-$direction). These domains are one lattice unit wide along the transverse direction ($y-$direction) and are independent of the neighboring rings [Fig.~\ref{fig:D_zero_snap}(b)]. As particles can not diffuse along the transverse direction, these domains can only grow horizontally without merging with the neighboring rings. At late times, we observe narrow horizontal stripes of alternating magnetization [Fig.~\ref{fig:D_zero_snap}(c)]. Therefore, AIM with $D_{\perp}=0$ does not manifest the observed domain morphology representative of the growth law $\sim t^{1/2}$. $D_{\perp}=0$ also signifies $L_y$ numbers of one-dimensional periodic rings on which the AIM is defined. Such $1d$ AIM has been found to display flocking of a single dense ordered aggregate at intermediate temperatures (this flock undergoes stochastic reversals of its magnetization with time) but an aster phase consisting of sharp peaks of positive and negative magnetizations in a jammed state at lower temperatures~\cite{benvegnen2022flocking}. The one lattice unit-wide (along $y$) domains in Fig.~\ref{fig:D_zero_snap}(c) manifest the characteristics of the flocking state of $1d$ AIM for the given parameters (see Appendix~\ref{appB} for details) but the system does not exhibit flocking as a whole since the $L_y$ number of $1d$ rings do not interact with each other for $D_{\perp}=0$. Altering the transverse diffusion $D_{\perp}$ to a nonzero value, particle diffusion occurs along the transverse $\pm y$ direction. Therefore, the random initial state coarsens to form high-density clusters which coarsen further to give rise to a large flocking domain [Fig.~\ref{fig:D_zero_snap}(d-f)]. 

\begin{figure}[!t]
    \includegraphics[width=\columnwidth]{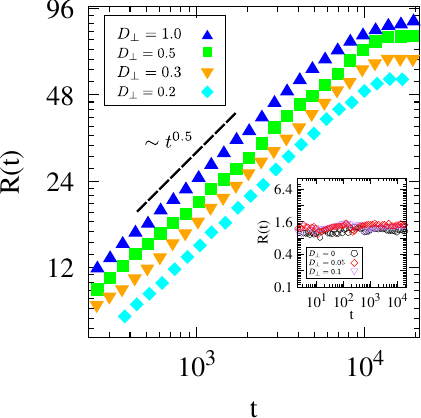}
    \caption{(Color online) $R(t)$ versus $t$ (on a log-log scale) showing a 0.5 growth exponent for different transverse diffusion $D_{\perp}= 0.2$ (solid cyan diamond), 0.3 (solid yellow inverted triangle), 0.5 (solid green square), and 1 (solid blue triangle). Inset: $R(t)$ versus $t$ for small values of the transverse diffusion, $D_{\perp}=0$ (open black inverted pentagon), 0.05 (open red diamond), and 0.1 (open purple inverted triangle) showing no temporal domain growth. Parameters: $L=400$, $\beta=1.1$, $\epsilon=1$, and $\rho_0=8$.}
    \label{fig:diff_D}
\end{figure}

To quantify the role of the diffusion coefficient ($D_{\perp}$), we plot $R(t)$ versus $t$ in Fig.~\ref{fig:diff_D} for various values of $D_{\perp}$. The plot shows that the system follows a power-law growth of $R(t)\sim t^{1/2}$ across different values of $D_{\perp}$. Notably, even a small value of $D_{\perp}=0.2$ is sufficient to drive proper domain growth. Increasing $D_{\perp}$ further only increases the average domain size. As shown earlier in Fig.~\ref{fig:D_zero_snap}(b--c), a vanishing $D_{\perp}$ or a very small $D_{\perp}$ can not initiate domain growth, as $R(t)$ remains constant with $t$ (see inset of Fig.~\ref{fig:diff_D}). This indicates that transverse diffusion plays one of the most crucial roles in the growth kinetics of the AIM. However, our primary motivation for suppressing transverse diffusion extends beyond simply verifying the diffusion-driven nature of coarsening. As shown in Fig.~\ref{fig:diff_D}, we explore the critical regime of transverse diffusion ($D_{\perp}$) where the growth exponent transitions to the diffusive behavior observed in Model A. By systematically reducing $D_{\perp}$ and analyzing the resulting growth dynamics, we identify the limiting values of $D_{\perp}$ at which this transition occurs. This investigation provides insights into the sensitivity of the growth kinetics of the AIM to the strength of transverse diffusion.


\subsection{Growth law from the hydrodynamic description of the AIM}
\label{hydro}
In this section, we want to investigate whether the continuous description of the {\color{blue}AIM~\cite{AIM_solon_intro,solon2015flocking,kourbane2018exact,scandolo2023active}} also manifests the same time dependency $(\sim t^{1/2})$ of the coarsening length scale as observed in our numerical analysis. We consider {\it refined} mean-field equations similar to Ref.~\cite{solon2015flocking} for the spatiotemporal evolutions of the density $(\rho)$ and magnetization $(m)$ fields:
\begin{gather}
    \Dot{\rho}=D\nabla^2\rho-v\partial_x m \, , \label{rmf1} \\ 
    \Dot{m}=D\nabla^2 m-v\partial_x \rho + 2\left(2\beta-1-\frac{r}{\rho}\right)m-\alpha\frac{m^3}{\rho^2} \, , \label{rmf2}
\end{gather}
where $v=2D\epsilon$, $\alpha=4\beta^2(1-2\beta/3)$, and $r=3\alpha\alpha_m/2$ is a positive function of $\beta$~\cite{solon2015flocking}. In constructing Eq.~\eqref{rmf2}, we slightly modify the flipping rate equation of Eq.~\eqref{flip} to read $W_{\rm flip}=\exp\left(-2\beta\sigma m/\rho\right)$. For the mean-field equations, $r = 0$, which cannot correctly capture AIM physics as the system always exhibits homogeneous profiles (gas and liquid), and inhomogeneous phase-separated profiles are never observed~\cite{solon2015flocking}. In Eqs.~\eqref{rmf1} and \eqref{rmf2}, local fluctuations are taken into account, which are generally neglected in the mean-field approximations~\cite{solon2015flocking}.
\begin{figure}[!t]
    \includegraphics[width=\columnwidth]{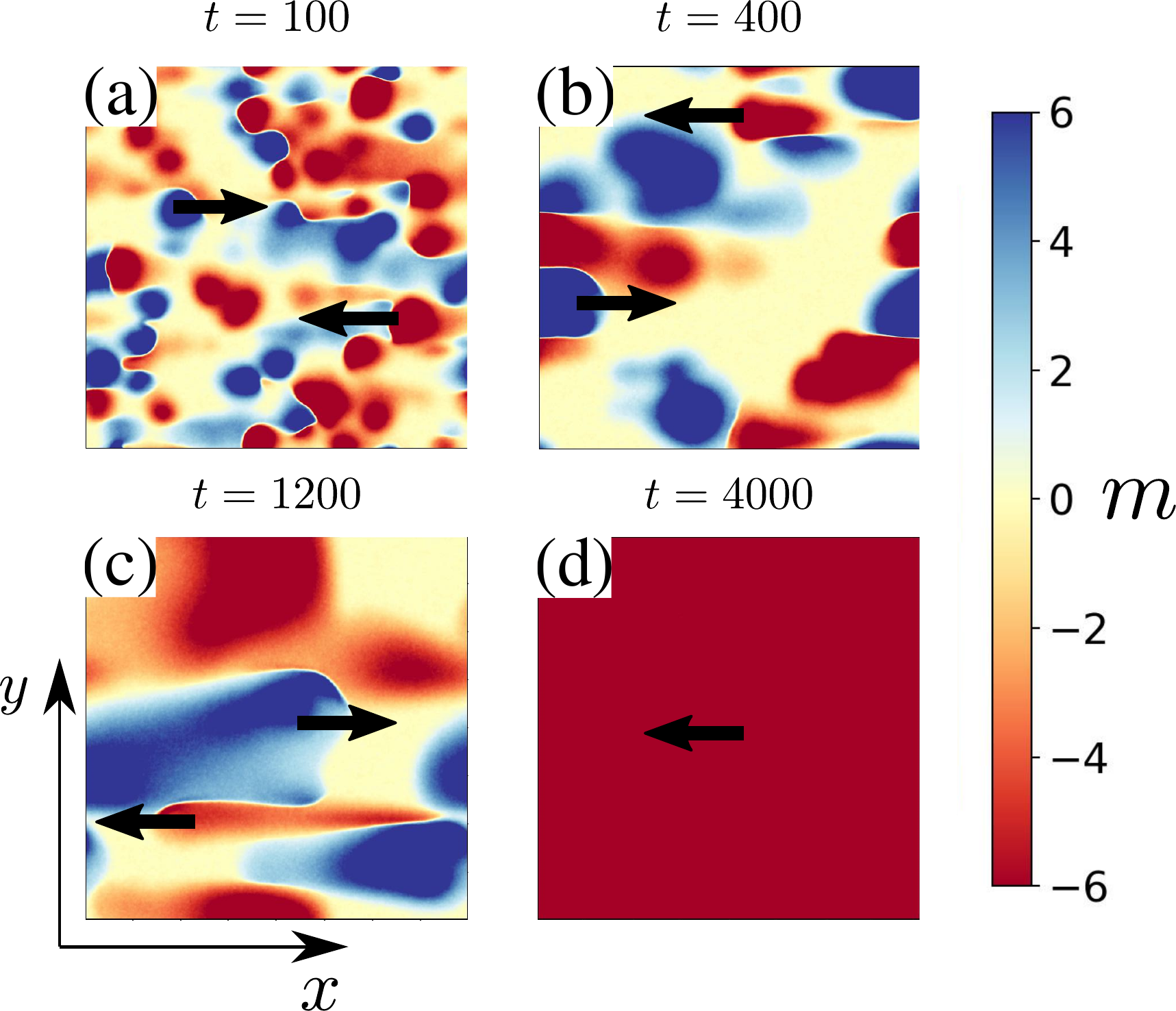}
    \caption{(Color online) Time evolution snapshots of the magnetization field $m$ obtained by solving the AIM hydrodynamic equations Eq.~\eqref{rmf1} and Eq.~\eqref{rmf2} after a quench from the homogeneous gaseous phase to an ordered liquid phase. Parameters: $L=400$, $\beta=0.75$, and $\rho_0=5$.}
    \label{fig:Snap_Hydro}
\end{figure}

Eqs.~\eqref{rmf1} and \eqref{rmf2} immediately show the importance of transverse diffusion. If we start with a $x$-independent initial condition, for instance, a horizontal thin stripe of width $w$, where: $\rho(y)=\rho_0/w$, $m(y)=m_0>0$ is the stationary solution of Eq.~\eqref{rmf2} for $0<y<w$, and $\rho(y)=\delta \ll \rho_0/w$ and $m_0=0$ otherwise. Then the solution stays $x$-independent at all later times: $\rho(x,y,t)=\Tilde{\rho}(y,t)$ and $m(x,y,t)=\Tilde{m}(y,t)$ where the equations for $\tilde \rho$ and $\tilde m$ are:
\begin{gather}
    \Dot{\Tilde{\rho}}=D\partial_{yy}\Tilde{\rho} \, , \label{rmf5} \\ 
    \Dot{\Tilde{m}}=D\partial_{yy} \Tilde{m} + 2\left(2\beta-1-\frac{r}{\Tilde{\rho}}\right)\Tilde{m}-\alpha\frac{\Tilde{m}^3}{\Tilde{\rho}^2} \, . \label{rmf6}
\end{gather}
Eq.~\eqref{rmf5} shows that the thin stripe expands diffusively in the $y$-direction (transverse direction) yielding a $\sqrt{t}$ dependence of the stripe width. For general random initial conditions, we use explicit Euler FTCS (Forward Time Centered Space)~\cite{press2007numerical} differencing scheme to numerically integrate Eqs.~\eqref{rmf1} and \eqref{rmf2}. We solve these two coupled partial differential equations on a square domain of size $L \times L$ with periodic boundary conditions applied in both directions. In our simulation, $L=400$ and the maximum simulation time is $t_{\rm sim}=5 \times 10^6$. To maintain the numerical stability criteria, we set $\Delta x = 1$ as the discretization in space and $\Delta t = 10^{-3}$ as the discretization in time. These discretization parameters satisfy the Courant-Friedrichs-Lewy (CFL) stability condition. In our numerical implementation, we fix $D=r=v=1$ and the initial system is prepared as a high-noise homogeneous gas phase with $\rho=\rho_0$ and $m=0$ by adding a zero-mean scalar Gaussian white noise to Eq.~\eqref{rmf2}~\cite{solonVM}. We then calculate the spatial dependence of the density and magnetization correlation using Eqs.~\eqref{correlation_density} and \eqref{correlation_mag} for 25 independent realizations. Finally, $R(t)$ is determined where the ensemble-averaged correlation functions decay to 0.2 of its maximum value. 
\begin{figure}[!t]
    \includegraphics[width=\columnwidth]{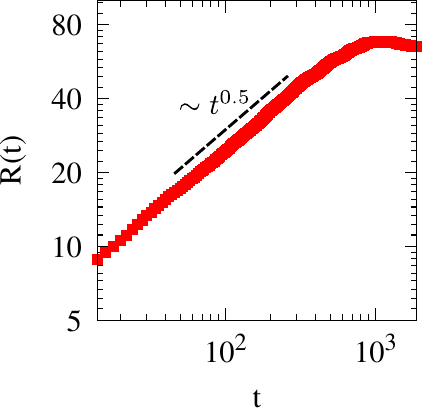}
    \caption{(Color online) $R(t)$ versus $t$ (on a log-log scale) exhibiting a growth exponent $\theta=1/2$ by solving the AIM hydrodynamic equations Eq.~\eqref{rmf1} and Eq.~\eqref{rmf2} using the FTCS scheme. Parameters: $L=400$, $\beta=0.75$, and $\rho_0=5$.}
    \label{fig:R_Hydro}
\end{figure}

In Fig.~\ref{fig:Snap_Hydro}, we plot the time evolution of the magnetization field $m$ by solving Eqs.~\eqref{rmf1} and \eqref{rmf2} via the finite difference FTCS scheme. The formation of self-propelling clusters with smooth interfaces and their growth with time resembles the dynamics shown in Fig.~\ref{fig:dynamics_snap} for the time-evolution of the microscopic model [Eqs.~(\ref{hamiltonian}--\ref{hop})]. We also extract the LCA growth law $R(t) \sim t^{1/2}$ (as shown in Fig.~\ref{fig:R_Hydro} on a logarithmic scale) by solving the hydrodynamic equations after quenching the system from a disordered gaseous phase to an ordered liquid phase. The length scale in Fig.~\ref{fig:R_Hydro} is obtained from the equal-time spatial correlation of the density fields (the length scale obtained from the correlation of the magnetization fields also exhibits the same growth law). Hence, the self-propulsion terms in the hydrodynamic equations \eqref{rmf1} and \eqref{rmf2} of the AIM do not alter the asymptotic growth law of the non-conserved Model A.

It can be seen that for $v=0$ $(\epsilon=0)$, Eq.~\eqref{rmf2} transforms into the deterministic version of the well-known time-dependent Ginzburg–Landau (TDGL) equation~\cite{sanjay_puri} of the order parameter evolution. Then one can derive the Allen-Cahn equation of motion for the interfaces from the TDGL equation~\cite{sanjay_puri}, and this will produce a growth $R(t) \sim t^{1/2}$ as shown in Fig.~\ref{fig:growth_plot}(c). Eq.~\eqref{rmf1} for $v=0$ follows a purely diffusive equation and therefore the density length scale will also exhibit a $\sqrt{t}$ dependence.

We now aim to discuss and compare the impact of hydrodynamics on domain growth in both the AIM and the VM. From Eqs.~\eqref{rmf1} and \eqref{rmf2}, we observe that in the AIM, density is advected by the magnetization only in the $x$-direction and vice versa. Therefore, diffusion, denoted by the term $D\nabla^2 \rho$ (or $D\nabla^2 m$) becomes exceptionally significant in the growth of the density (magnetization) domains in the $2d$ AIM. Now, let us consider the minimal hydrodynamic equations that can describe the VM~\cite{solonVM,solon2015pattern}:
\begin{gather} 
    \partial_t\rho = - v \vec{\nabla} \cdot \vec{m} \, , \label{vmeq1} \\ 
    \partial_t \vec{m} = D\nabla^2 \vec{m} - \lambda \vec{\nabla}\rho - \xi \left(\vec{m} \cdot \vec{\nabla}\right) \vec{m} + \left(a_2 - a_4 | \vec{m} |^2 \right) \vec{m} \, , \label{vmeq2}
\end{gather}
where $\vec{m}(\vec{r},t)$ is the vectorial magnetization field for continuous symmetry, $a_2$ and $a_4$ are functions of $\rho$, $\lambda\vec{\nabla}\rho$ reflects the pressure gradient induced by density heterogeneities, $\xi$ and $D$ are two transport coefficients associated with the advection and the diffusion of the magnetization, respectively. As one can notice, the advective term $v \partial_x m$ of the AIM is replaced by $v \vec{\nabla} \cdot \vec{m}$ in Eq.~\eqref{vmeq1}, and in the VM, density can only be advected along the direction of the magnetization $\vec{m}$. As there is no $D\nabla^2\rho$ term in the Vicsek density equation, diffusion cannot play a role in the growth of the density domains by diffusing particles in directions other than those dictated by $\vec{m}$. Ordering dynamics in self-propelled particles with the diffusive term in the density equation has been reported to show a faster growth of the density field~\cite{das2018ordering}. In Eq.~\eqref{vmeq2}, along with the diffusive term $D\nabla^2 \vec{m}$, the additional $\left(\vec{m} \cdot \vec{\nabla}\right) \vec{m}$ term is an advective term (analogous to the advective term in the Navier-Stokes equation). It is well known that in Model H~\cite{hohenberg1977}, where advection is included (a conserved order parameter is coupled to hydrodynamic flow), the asymptotic growth exponent is greater than 0.5: $R(t) \sim t$ when the advective term is dominant at late times (viscous hydrodynamic regime), and even later times, when the inertial effect also becomes important (inertial hydrodynamic regime), $R(t) \sim t^{0.67}$~\cite{bray2003coarsening}. Therefore, advection is probably most relevant for the emergence of the large value (0.83) in the growth exponent for magnetization in the VM. In contrast to the VM, within the AIM, the {\it flow} occurs exclusively along the $x$ direction, emphasizing the crucial role of diffusion in growth kinetics. As a result, both density and magnetization grow diffusively.


\section{Summary and Discussion}
\label{conclusion}
We conclude this paper with a summary and discussion of our results. We study the ordering kinetics of the active Ising model (AIM), a flocking model with both conserved (density) and non-conserved (magnetization) scalar order parameters, after quenching from a disordered high-temperature gaseous phase to the phase coexistence region and the polar-ordered liquid phase. We observe the formation of connected domains of negative and positive spins similar to the ferromagnetic Ising model in the non-motile diffusive limit of the AIM. But for self-propelled particles, AIM manifests an extensive number of disconnected small clusters of the negative and positive spins which eventually merge to a single, macroscopic liquid domain~\cite{solon2015flocking}. The domain evolution morphology is characterized by the equal-time two-point correlation function and its Fourier transform, the structure factor. The scaling behavior of the correlation function demonstrates a good data collapse, indicating the self-similar nature of domain growth. Additionally, the large-$k$ behavior of the scaled structure factor tail exhibits Porod's decay, supporting the smooth spatial structure of the AIM domains. Although AIM contains both conserved and non-conserved order parameters, we extract the same growth law for the AIM density and magnetization fields, the Lifshitz-Cahn-Allen (LCA) growth law~\cite{alan_bray,bray1993theory,sanjay_puri} $R(t)\sim t^{1/2}$ of the non-conserved scalar order parameter. This indicates that activity does not affect the growth law of the AIM. This observation aligns with earlier findings in Model A with an external drive where external influences do not alter the growth law at the linear level~\cite{bray2001interface}. Unlike the VM~\cite{Coarsening2020VM}, in AIM, the density domain aligns over the same length scale as the orientation. Furthermore, we do not observe any activity-induced correction to the growth law of the non-conserved scalar order parameter, as observed in the context of the active polar fluid with a non-conserved vector order parameter~\cite{Dikshit_2023}. Further, we investigate the role of diffusion in the growth kinetics of the AIM. We observe that due to the horizontal biased hopping (along $\pm x$-direction) of the AIM clusters, particle diffusion along the vertical $\pm y$-direction is the predominant mechanism through which coarsening happens in the AIM.  This establishes diffusion as the main growth mechanism rather than activity, leading to a growth law $R(t)\sim t^{1/2}$ for the AIM. We further solve the AIM hydrodynamic equations using a finite difference scheme, and the extracted coarsening length scale validates the growth exponent $\theta \sim 1/2$ observed in the microscopic simulation. As a future exercise, one could also consider exploring an analytical approach~\cite{bray2001interface} to solve the hydrodynamic equations to perceive a better understanding of the coarsening dynamics. 

In this paper, we focus on purely local particle interactions. Instead of considering a purely local interaction between particles, one can also consider the presence of interactions between the nearest neighboring sites and investigate whether this short-range spatial structure of the interactions affects the extracted growth law. Upon coarse-graining, coupling the magnetization on a site to those on its four neighboring sites will only modify the magnetization equation, but not the density equation, which will retain its form. Consequently, we expect the same growth law exponent in the presence of nearest-neighbor interactions.

An interesting extension of this study would be to explore the phase ordering kinetics in flocking models in the presence of disorder, as experimental systems always contain both quenched and mobile impurities. The similarity in the growth law of a passive system and its active counterpart is an interesting result. Therefore, further studies on the coarsening dynamics of active systems, such as the active Potts model or the active clock model, where the growth laws of the corresponding passive models are well known~\cite{grest1988domain,chatterjee2018clock,rbcm}, are required to confirm (or contradict) this theoretical observation. 

\section{Acknowledgments}
MK acknowledges financial support through a research fellowship from CSIR, Govt. of India (Award Number: 09/080(1106)/2019-EMR-I). RP, SB, AD, and MK thank the Indian Association for the Cultivation of Science (IACS) for its computational facility and resources. SC and HR are financially supported by the German Research Foundation (DFG) within the Collaborative Research Center SFB 1027-A3 and INST 256/539-1. SC acknowledges many helpful discussions with Dr Matthieu Mangeat.

\appendix
\section{Comparison of the growth law for the density and magnetization field}
\label{appA}
\begin{figure}
    \centering
    \includegraphics[width=\columnwidth]{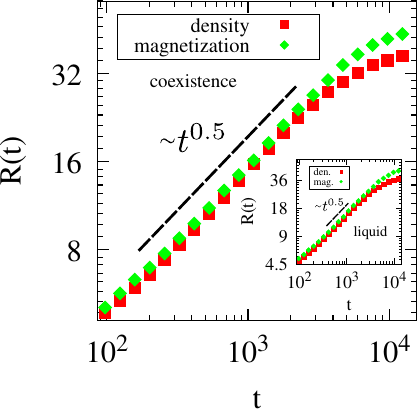}
    \caption{(Color online) $R(t)$ versus $t$ for density (solid red square) and magnetization (solid green diamond) field yields the same growth exponent 0.5 for a quench in the coexistence regime, $\rho_0=3$. (Inset) A similar growth law is extracted for a quench in the liquid regime, $\rho_0=10$. Parameters: $L=400$, $\beta=1.1$ and $\epsilon=1$.}
    \label{fig:mag_vs_den_compare}
\end{figure}
The ordering kinetics of the AIM discussed in this paper are studied mainly by extracting the length scale $R(t)$ from the equal-time two-point density correlation function defined in Eq.~\eqref{correlation_density}. However, one can also extract $R(t)$ from the magnetization correlation function defined in Eq.~\eqref{correlation_mag}. Now, it was argued in the context of the coarsening dynamics of the VM that, unlike generic coarsening systems which typically exhibit a single dominant length scale, the Vicsek model exhibits distinct coarsening length scales for the density and velocity correlations \cite{Coarsening2020VM}. In VM, despite the density and velocity fields being fully coupled, the velocity length scale grows much faster compared to the density length scale because velocity order extends over longer distances than density clusters due to the irregular fractal morphology of the density clusters. In AIM, however, besides being the density and magnetization fields fully coupled, the clusters are also regularly shaped with smooth boundaries, and therefore, the temporal behavior of the two length scales are found similar as shown in Fig.~\ref{fig:mag_vs_den_compare}.

Fig.~\ref{fig:mag_vs_den_compare} shows, for the coarsening of the AIM, that the two length scales exhibit the same growth law, $R(t) \sim t^{1/2}$ (although the domain size for the density field is marginally larger than the corresponding magnetization field) for a quench into the coexistence region and into the polar ordered liquid regime (see inset of Fig.~\ref{fig:mag_vs_den_compare}). Therefore, we can conclude that the $R(t) \sim t^{1/2}$ growth law is reasonably universal in the AIM as it neither depends on the quenching regime nor the local order parameter (be it conserved density or non-conserved magnetization). 

\section{2\textit{d} AIM with \texorpdfstring{\bm{$D_\perp=0$}}{Lg}}
\label{appB}
\begin{figure}
    \centering
    \includegraphics[width=\columnwidth]{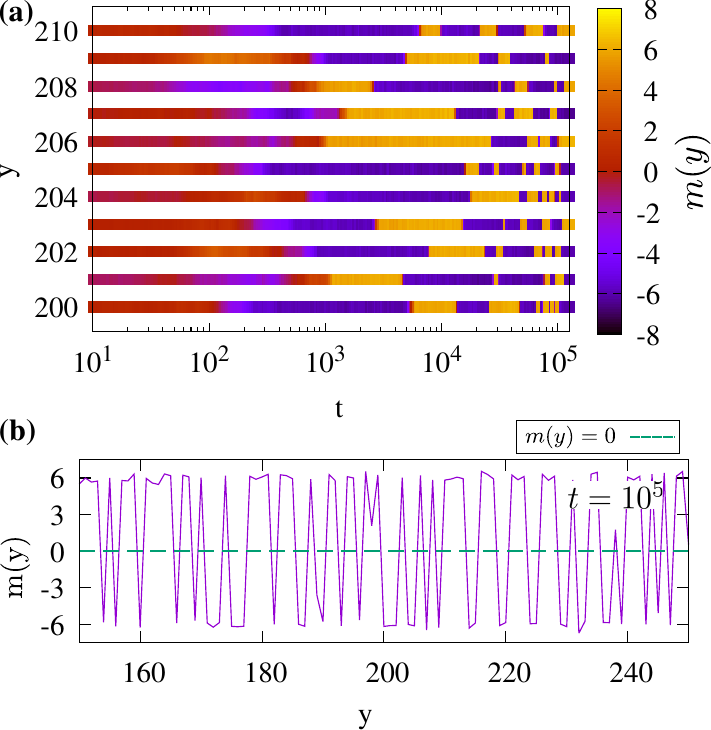}
    \caption{(Color online) (a) $m(y)$ versus $t$ for $L_y\in[200:210]$ starting from a random disordered configuration ($m \sim 0$). The color bar denotes the averaged magnetization along the $x$-axis. (b) Late time averaged magnetization profile as a function of $y$. Parameters: $L=400$, $\rho_0=10$, $\beta=1.1$, and $\epsilon=0.8$.}
    \label{fig:mag_D_zero}
\end{figure}   
The role of transverse diffusion in domain growth of $2d$ AIM is immense. Here we present a more detailed picture of the $D_\perp=0$ scenario. In Fig.~\ref{fig:mag_D_zero}(a) we plot $m(y)= \frac{1}{L_x}\sum^{L_x}_{i=1} m(i,y)$ versus time $t$ for $L_y\in[200:210]$. As the density is very large $(\rho_0=10)$, we see highly magnetized one lattice unit-wide domains that can not merge to create a larger domain as time progresses because we inhibit the diffusive hopping in the transverse direction. These $1d$ domains are single dense ordered aggregates that stochastically reverse their magnetization~\cite{benvegnen2022flocking}. The magnetization profile in Fig.~\ref{fig:mag_D_zero}(b) is averaged over $x$ and shows sharp peaks of positive and negative magnetizations, spread over either one site or a few sites. The density profile of such an arrangement shows a homogeneous profile around the average density which is similar to the density profile of a large liquid domain but the alternating magnetization profile signifies that there is no domain growth at the asymptotic limit in the $2d$ AIM for $D_{\perp}=0$. 

\bibliography{referenceaim}
\end{document}